\pdfoutput=1
\documentclass[preprint]{revtex4-1} 
\usepackage{graphicx}  
\usepackage{amsfonts}
\usepackage{geometry}
\usepackage[frozencache=true]{minted}
\usepackage[T1]{fontenc}
\usepackage{lmodern}
\usepackage{caption}

\begin{document}
\title{A Practical Python API for Querying AFLOWLIB}
\author{Conrad W. Rosenbrock}
\affiliation{Department of Physics and Astronomy, Brigham Young
  University, Provo, Utah 84602, USA.}
\date{\today}

\begin{abstract}
  Large databases such as \verb|aflowlib.org| provide valuable data
  sources for discovering material trends through machine
  learning. Although a REST API and query language are available,
  there is a learning curve associated with the AFLUX language that
  acts as a barrier for new users. Additionally, the data is stored
  using non-standard serialization formats. Here we present a
  high-level API that allows immediate access to the \verb|aflowlib|
  data using standard python operators and language features. It
  provides an easy way to integrate \verb|aflowlib| data with other
  python materials packages such as \verb|ase| and \verb|quippy|, and
  provides automatic deserialization into \verb|numpy| arrays and
  python objects. This package is available via \verb|pip install aflow|.
\end{abstract}

\maketitle

\section{Introduction}

Recent advances in computation have enabled the creation of large,
materials databases using Density Functional Theory
\cite{PhysRev.140.A1133,PhysRev.136.B864}. \verb|aflowlib|~\cite{Curtarolo2012218,Curtarolo2012227}
is one of the largest with more than 1.7M material compounds (as of
September 2017, http://aflowlib.org/). Recently, the AFLUX search API
\cite{ROSE2017362} was introduced to provide improved access to the
data in a uniform request format via REST
\cite{TAYLOR2014178}. Because the API is based on REST, it allows
access to the data from a variety of programming languages through
standard libraries.

Unfortunately, the data for material properties and calculation
parameters are not stored in a standard format. While the custom
serialization format is documented, each property must be parsed
individually to access standard formats (such as \verb|numpy|
\cite{5725236} arrays for python). Thus, a researcher attempting to
access \verb|aflowlib| data for the first time must 1) read and
understand the AFLUX request format; 2) lookup the documentation for
the properties of interest and 3) deserialize them
appropriately. \verb|aflowlib| fields are stored as strings of values
that may be comma-separated, colon-separated or have a more complex
structure (such as for the \verb|kpoints| property). Deserialization refers
to the transformation of these strings into high-level objects such as
dictionaries or arrays. Even though such tasks are well within the
abilities of a computational scientist, they are not tasks that
leverage scientific expertise.

Here, we introduce a high-level python API that abstracts the request
and deserialization tasks so that there is virtually no access barrier
for newcomers to the \verb|aflowlib| database.

\section{Request Example}

We begin with the example from the AFLUX paper \cite{ROSE2017362}, translated into python
using the new package (\verb|aflow|, https://pypi.python.org/pypi/aflow):

\begin{listing}[H]
\begin{minted}[mathescape,
               linenos,
               numbersep=5pt,
               frame=lines,
               framesep=2mm]{python}
from aflow import *

# Iterate over the results in batches of 20; sort by thermal
# conductivity descending.
result = search(batch_size=20
        ).select(K.agl_thermal_conductivity_300K
        ).filter(K.Egap > 6
        ).orderby(K.agl_thermal_conductivity_300K, True)

# Now, you can just iterate over the results.
for entry in result:
    print(entry.Egap, entry.agl_thermal_conductivity_300K)
\end{minted}
 \label{lst:aflux_search_example}
  \caption{Example of searching the aflowlib database for
    materials with large band gaps and large thermal
    conductivity. This is the same example given in the AFLUX paper \cite{ROSE2017362}.}
\end{listing}

The \verb|search| function returns an object that can be chained
continually to apply multiple filters (\verb|filter|), select
additional properties (\verb|select|), apply exclusions
(\verb|exclude|) or order the results
(\verb|orderby|). \verb|aflowlib| provides more than 110 keywords for
various material properties, calculation parameters, etc. Somebody new
to \verb|aflowlib| may not initially know what is available and would
have to pull up the documentation online. For ease-of-use, the API
provides all keywords supported by \verb|aflowlib| as attributes of
\verb|aflow.K|. For IDEs that support auto-complete, this means that
researchers can dynamically see what properties are available and view
descriptions of them by typing \verb|K.<tab>| and using tab completion.

\section{API Features}

The \verb|search| function described above produces an object that
supports iteration over the database entries returned from
\verb|aflowlib|. Normally, AFLUX requires desired properties to be
specified as part of the request URL. Our python API provides ``lazy
evaluation'' functionality so that database entries from an existing
result can be queried for additional properties using attributes on
the python database entry objects. All requests happen transparently
in the background and are cached to optimize performance.

\subsection{Slicing, Indexing and Deserialization}
\begin{listing}[H]
\begin{minted}[mathescape,
               linenos,
               numbersep=5pt,
               frame=lines,
               framesep=2mm]{python}
# The API also allows for slicing of result sets
part = result[21:25]

# If a page of results isn't available in an existing batch, the next
# one is automatically retrieved from the server. Since our batch size
# is 20, asking for result 55 triggers another paging request against
# the aflowlib API.
result[55] # <aflow.entries.Entry at 0x1104f9890>

# Our original query didn't ask for the atomic positions, but we can
# request them now anyway because of the lazy request evaluation.
result[55].positions_cartesian
#array([[ 0.    ,  0.    , -0.    ],
#       [ 1.5691,  1.5691,  1.5691],
#       [ 4.7073,  4.7073,  4.7073]])
\end{minted}
 \label{lst:aflow_features}
  \caption{The python API supports arbitrary slicing in the result set
    and lazy evaluation of properties. This means that properties can
    be fetched from the aflowlib database even if they weren't part of
    the original request URL.}
\end{listing}

Notice that the Cartesian positions are returned as a \verb|numpy|
array automatically because the API handles deserialization.

\subsection{Operators}

The \verb|filter| method of the search object filters results in
\verb|aflowlib| using standard operators. We overloaded these
operators in python to provide an intuitive interface. These are
briefly described here:

\begin{enumerate}
\item\label{item:1} > and < behave as expected. However, these
  are overloaded for string comparisons in the spirit of the AFLUX
  endpoint. For example \verb|author < "curtarolo"| will match ``*curtarolo''
  and \verb|author > "curtarolo"| will match ``curtarolo*''.
\item\label{item:2} == behaves as expected for all keywords.
\item\label{item:3} \% allows for string searches.
  \verb|author % "curtarolo"| matches ``*curtarolo*''.
\item\label{item:4} \textasciitilde \, inverts the filter (equivalent to a boolean
  ``not'').
\item\label{item:5} \& is the logical ``and'' between two
  conditions.
\item\label{item:6} | is the logical ``or'' between two
  conditions.
\end{enumerate}

Using these operators, it is possible to form complex queries using
intuitive notation:

\begin{listing}[H]
\begin{minted}[mathescape,
               linenos,
               numbersep=5pt,
               frame=lines,
               framesep=2mm]{python}
# Find entries with a band gap between 0 and 2 or 5 and 7. 
filter((K.Egap > 0) & (K.Egap < 2)) | ((K.Egap > 5) & (K.Egap < 7))
\end{minted}
 \label{lst:complex_query}
 \caption{Example of chaining complex filters using the overloaded
   operators in the API. Because we overload the bit-wise \& and |,
   extra parentheses are required around the numeric operator
   expressions.}
\end{listing}

\subsection{Templated Generation}

The supported keywords and corresponding properties on the database
entry objects are generated via requests to the \verb|aflowlib|
schema, which includes documentation. This allows any additions or
modifications at \verb|aflowlib| to be captured automatically by
regenerating the python API from template.

\subsection{Integration with ASE and quippy}

The Atomic Simulation Environment (ASE)
\cite{ase-paper,ISI:000175131400009} provides a high-level API for
working with materials, calculating their properties and performing
other high-level transformations. \verb|quippy| (\cite{quippy},
http://www.libatoms.org) extends the functionality of \verb|ase| with
additional routines that make it easier to work with collections of
material configurations and perform additional tasks (such as crack
propagation simulations, calculating descriptors, etc.).

Each database entry object in the \verb|aflow| API also provides an
\verb|atoms| method that constructs an atoms object for \verb|ase| or
\verb|quippy|. This makes it seamless to integrate the \verb|aflowlib|
data into existing workflows.

\section{Installation and API Documentation}

The package is available on the python package index and can be
installed with: 

\verb|pip install aflow|

API Documentation auto-generated by sphinx is available at:

\verb|https://rosenbrockc.github.io/aflow/|

The source code has full test coverage and continuous integration, and
is hosted at

\verb|https://github.com/rosenbrockc/aflow|

\verb|aflow| works in both python 2 and python 3.

\end{document}